%% file: conference_101719.tex
\documentclass[conference]{IEEEtran}
\usepackage{amsmath,amssymb,amsfonts}
\usepackage{algorithmic}
\usepackage{graphicx}
\usepackage{textcomp}
\usepackage{xcolor}
\usepackage{orcidlink}
\usepackage{enumitem}
\setlist[itemize]{leftmargin=*}
\setlist[enumerate]{leftmargin=*}
\allowdisplaybreaks
\usepackage{xurl}
\usepackage{microtype}

\usepackage{subcaption}
\usepackage{url}
\usepackage{multirow}
\usepackage{array}
\usepackage[numbers]{natbib}
\usepackage{algorithmic}
\usepackage{algorithm}

\usepackage{tikz}
\usepackage{textcomp}
\usepackage{hyperref}
\makeatletter
\pretocmd{\NAT@citexnum}{\@ifnum{\NAT@ctype>\z@}{\let\NAT@hyper@\relax}{}}{}{}
\makeatother

\def\BibTeX{{\rm B\kern-.05em{\sc i\kern-.025em b}\kern-.08em
    T\kern-.1667em\lower.7ex\hbox{E}\kern-.125emX}}
\begin{document}

\title{Ponder: Online Prediction of Task Memory Requirements for Scientific Workflows}

\author{\IEEEauthorblockN{Fabian Lehmann\textsuperscript{\orcidlink{0000-0003-0520-0792}}$^{1}$, Jonathan Bader\textsuperscript{\orcidlink{0000-0003-0391-728X}}$^{2}$, Ninon De Mecquenem\textsuperscript{\orcidlink{0000-0003-3052-6129}}$^{1}$,\\ Xing Wang\textsuperscript{\orcidlink{0000-0001-8645-9354}}$^{1}$, Vasilis Bountris\textsuperscript{\orcidlink{0000-0002-8682-7302}}$^{1}$, Florian Friederici\textsuperscript{\orcidlink{0009-0005-0688-696X}}$^{1}$, Ulf Leser\textsuperscript{\orcidlink{0000-0003-2166-9582}}$^{1}$, Lauritz Thamsen\textsuperscript{\orcidlink{0000-0003-3755-1503}}$^{3}$}
\IEEEauthorblockA{\small
\textit{$^1$Humboldt-Universität zu Berlin, Germany,}\\
\textit{\{fabian.lehmann, mecquenn, xing.wang, vasilis.bountris, friederf, leser\}@informatik.hu-berlin.de}\\
\textit{$^2$Technische Universität Berlin, Germany, jonathan.bader@tu-berlin.de}\\
\textit{$^3$University of Glasgow, United Kingdom, lauritz.thamsen@glasgow.ac.uk}}}

\newcommand\copyrighttext{%
    \footnotesize \textcopyright 2024 IEEE. Personal use of this material is permitted.
    Permission from IEEE must be obtained for all other uses, in any current or future
    media, including reprinting/republishing this material for advertising or promotional
    purposes, creating new collective works, for resale or redistribution to servers or
    lists, or reuse of any copyrighted component of this work in other works.
    DOI: \href{https://doi.org/10.1109/e-Science62913.2024.10678682}{https://doi.org/10.1109/e-Science62913.2024.10678682}}
\newcommand\copyrightnotice{%
    \begin{tikzpicture}[remember picture,overlay]
        \node[anchor=south,yshift=10pt] at (current page.south) {\fbox{\parbox{\dimexpr\textwidth-\fboxsep-\fboxrule\relax}{\copyrighttext}}};
    \end{tikzpicture}%
}

\maketitle
\copyrightnotice

\begin{abstract}
\input{content/00_abstract}
\end{abstract}


\input{content/01_intro}

\input{content/02_problemanalysis}
\input{content/02.5_executionanalysis}

\input{content/03_strategy}
\input{content/05_experiments}

\input{content/06_relatedwork}
\input{content/99_conclusion}

\section*{Acknowledgment}

This work was funded by the Deutsche Forschungsgemeinschaft (DFG, German Research Foundation) – Project-ID 414984028 – SFB 1404 FONDA.

\bibliographystyle{IEEEtranN}
\bibliography{bib}

\end{document}

%% file: content/00_abstract.tex
Scientific workflows are used to analyze large amounts of data.
These workflows comprise numerous tasks, many of which are executed repeatedly, running the same custom program on different inputs.
Users specify resource allocations for each task, which must be sufficient for all inputs to prevent task failures.
As a result, task memory allocations tend to be overly conservative, wasting precious cluster resources, limiting overall parallelism, and increasing workflow makespan.

In this paper, we first benchmark a state-of-the-art method on four real-life workflows from the nf-core workflow repository.
This analysis reveals that certain assumptions underlying current prediction methods, which typically were evaluated only on simulated workflows, cannot generally be confirmed for real workflows and executions.
We then present \emph{Ponder}, a new online task-sizing strategy that considers and chooses between different methods to cater to different memory demand patterns.
We implemented Ponder for Nextflow and made the code publicly available.
In an experimental evaluation that also considers the impact of memory predictions on scheduling, Ponder improves Memory Allocation Quality on average by 71.0\% and makespan by 21.8\% in comparison to a state-of-the-art method.
Moreover, Ponder produces 93.8\% fewer task failures.\looseness=-1

%% file: content/01_intro.tex
\section{Introduction}
Scientific workflows are becoming increasingly important in many fields, including remote sensing~\cite{WilkinsonEnvironmental2024, phiriSentinel22020}, bioinformatics~\cite{bioinfo_introduction,nf_core}, and astronomy~\cite{ahmadEfficient2022, berrimanTerapixel2022}.
Workflows consist of multiple abstract tasks that integrate custom programs into reproducible processing pipelines.
These tasks serve as a blueprint for physical task instances.
Each physical instance executes a task's custom program on concrete input data.

Workflows are structured as Directed Acyclic Graphs (DAGs), where vertices represent tasks and edges represent dependencies between tasks.
Figure~\ref{fig:exampleAbstractDAG} shows an abstract workflow DAG with five different tasks.
Given concrete input data, the tasks are initialized, and a physical DAG is created, as shown in Figure~\ref{fig:exampleDAG}.
Notably, $c_1$ and $c_2$ are instances of the same abstract task $C$.
This is a typical case in which multiple physical tasks are instantiated based on the same abstract task, applying the same program to different data.

\input{figures/dagHori}

To execute a workflow, a Scientific Workflow Management System (SWMS) takes each physical task that is ready to run, meaning that all tasks it depends on have been completed, and sends it to a resource manager~\cite{dSFS+17, lehmannHowWorkflowEngines2023}.
For example, to submit $e_1$ in Figure~\ref{fig:exampleDAG}, the tasks $c_1$, $c_2$, and $d_1$ must have been finished.
The resource manager subsequently assigns tasks to nodes based on resource requirements and availability.
SWMSs provide the specification of the required resources, usually a certain number of cores and amount of main memory. 
This specification stems from workflow definitions, i.e., it is configured by a workflow developer based on (if available) past executions and educated guessing~\cite{wittLearningLowWastageMemory2019}.
At the same time, the resource requirements significantly impact workflow execution.
CPU underprediction will increase a task's runtime, while memory underprediction leads to a task's failure, as resource managers kill tasks that exceed their memory demands.
To avoid this, users resort to conservative estimates.
However, overprediction wastes resources, which then cannot be assigned to other tasks, limiting parallelism and increasing the workflow's makespans.\looseness=-1

Fine-grained user estimates for every physical task are hardly practical due to their sheer number.
Furthermore, the dynamic unfolding of workflows in modern SWMSs means that physical tasks may be determined only at runtime~\cite{buxHiWAYExecutionScientific2017}.
Hence, resource requirements are specified for abstract tasks.

There are two pitfalls to specifying resource requirements for abstract tasks.
First, a resource requirement for a task has to work for all physical instances of a given task, even though the resource demands of those instances might vary substantially depending on input size.
Therefore, resource requirements target maximal resource demands which introduces overhead for all but the largest instances and might exclude nodes with resources that are insufficient for the maximal, yet sufficient for most task instances.
Second, many workflows are being reused.
In these cases, the task resource requirements must work not only for current data but also for future data. %
Both pitfalls suggest resorting to coarse-grained, conservative estimates of memory requirements, and indeed, this is what is done in practice.
For example, the popular workflow collection nf-core~\cite{nf_core} simplifies task resource requirements and categorizes them into ``single-core'', ``low'', ``medium'', ``high'', and ``high-memory'' processes.
Tasks in the same category get the same resources, which inevitably leads to overprovisioning in many cases.\looseness=-1

Automated peak memory prediction approaches for workflow tasks have been developed to address the problem of coarse-grained, conservative, user-provided predictions~\cite{tovarJobSizingStrategy2018a,wittFeedbackBasedResourceAllocation2019,wittLearningLowWastageMemory2019,baderLeveragingReinforcementLearning2022,bader2023predicting}.
These approaches have the advantage that predictions are made for each physical task instance, decreasing memory wastage considerably.
The approaches can be categorized into methods that are trained offline on historic execution traces, online providing a newly updated model each time task executions finish, or in a hybrid manner where models can be pre-trained on historic execution traces and further updated at runtime.
Online approaches~\cite{tovarJobSizingStrategy2018a,wittFeedbackBasedResourceAllocation2019,baderLeveragingReinforcementLearning2022,bader2023predicting} are the most practical as they work out of the box without the need for any historic execution traces.
However, these approaches either rely on a linear correlation between input data size and the memory requirement of a task~\cite{wittFeedbackBasedResourceAllocation2019,wittLearningLowWastageMemory2019,bader2023predicting} or assume that previously seen peak memory demands will hold true also for future task instances~\cite{wittFeedbackBasedResourceAllocation2019,tovarJobSizingStrategy2018a,baderLeveragingReinforcementLearning2022}.
Furthermore, except for \cite{baderLeveragingReinforcementLearning2022}, these approaches have only been validated using simulations and are built upon several simplifying assumptions: They assume linear dependencies between inputs and memory consumption, normally distributed errors, and that memory underprediction predictably leads to task failures after a fixed fraction of the total runtime a task needs to finish successfully.

In this paper, we shed light on how well online memory prediction methods can work in the real world.
For this, we first analyze the problem of peak memory prediction for workflow tasks, showing memory demand patterns and variances that occur in popular Nextflow workflows.
Second, we develop \emph{Ponder}, a new online peak memory sizing strategy for workflow tasks based on the findings of our problem analysis.
Third, we implement our strategy directly into the nf-cws plugin for Nextflow and the Common Workflow Scheduler, so that it is available for Nextflow workflow executions\footnote{\label{git:nfcws}\url{https://github.com/CommonWorkflowScheduler/nf-cws}}\footnote{\label{git:cws}\url{https://github.com/CommonWorkflowScheduler/KubernetesScheduler}}.
Finally, we experimentally compare our approach to a state-of-the-art online approach, Witt-LR~\cite{wittFeedbackBasedResourceAllocation2019}, which uses linear regression to estimate memory requirements and adds an offset to bias towards overprediction.
The outcomes, alongside all traces and measurements, are accessible on GitHub\footnote{\label{git:experiments}\url{https://github.com/CommonWorkflowScheduler/MemorySizingExperiments}}.
Note that Witt-LR has only been evaluated using simulated workflow executions so far.
The evaluation is performed by running four nf-core workflows on an eight-node commodity cluster.
We further evaluate the impact on scheduling, assessing four scheduling strategies, two of them in two different versions.

%% file: figures/dagHori.tex
\begin{figure}[b!]
    \centering
    \begin{subfigure}{0.5\columnwidth}
      \centering
      \includegraphics[width=0.6\columnwidth,clip]{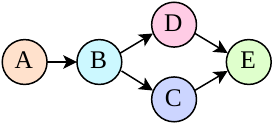}
      \caption{Abstract DAG}
	  \label{fig:exampleAbstractDAG}
    \end{subfigure}%
    \begin{subfigure}{0.5\columnwidth}
      \centering
      \includegraphics[width=0.6\columnwidth,clip]{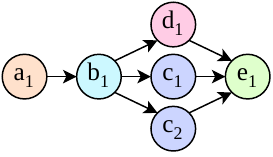}
      \caption{Physical DAG}
	  \label{fig:exampleDAG}
    \end{subfigure}%
    \caption{An abstract DAG with five tasks and a matching physical DAG with six physical tasks; circles are tasks, and arrows are dependencies.}
    \label{fig::abstractAndPhysicalDag}
\end{figure}

%% file: content/02_problemanalysis.tex
\section{Problem Analysis}
\label{sec:problem_analysis}
In this section, we analyze the problem of task peak memory demand prediction.
To this end, we first provide an overview of Nextflow and nf-core and then outline the workflows used for the experiments in this study.
We then empirically examine the memory consumption of tasks in these workflows and identify problems for automated memory prediction.

\subsection{Nextflow and nf-core}

Nextflow~\cite{ditommasoNextflowEnablesReproducible2017} is an SWMS that, in recent years, has received considerable attention\footnote{\url{https://www.nextflow.io/blog/2023/reflecting-on-ten-years-of-nextflow-awesomeness.html}}.
Nextflow addresses two important aspects of workflow development: portability and reproducibility. 
It is compatible with different execution environments and can be used on many different infrastructures.
Furthermore, Nextflow supports containerization and task execution via, for example, Docker or Singularity, allowing data analysis to be reproduced across execution environments.

nf-core is a community-curated collection of popular Nextflow workflows~\cite{nf_core}. 
It currently consists of 63 published workflows and 34 under development. 
The project set out to develop bioinformatics pipelines that would address the challenges of portability and reproducibility commonly encountered by this community.
Bioinformaticians frequently face difficulties in reproducing the outcomes of another study, a problem that is not exclusive to the field of bioinformatics. 
The set of nf-core workflows is growing, and workflows from other fields are currently under development. 
For example, there is meerpipe~\cite{Bailes2020_Meerkat}, an astronomy pipeline that processes pulsar data, or Rangeland~\cite{lehmannFORCENextflowScalable2021}, which analyzes satellite data.

nf-core workflows follow a set of best practices. 
Each nf-core workflow should be accompanied by extensive documentation on usage, input and output, and available parameters.
Furthermore, it should also include suitable test datasets for automated continuous integration tests. 
Therefore, the nf-core workflows are a valuable source of community-approved, actively maintained, and documented Nextflow workflows and are, therefore, ideal for testing, developing, and evaluating advanced workflow resource management methods.

\subsection{Workflows}\label{sec:workflows}
\input{figures/workflowInputs}

We choose the three most popular nf-core workflows measured by GitHub stars to test memory sizing approaches: RNA-Seq, Sarek~\cite{sarek_daw}, and MAG~\cite{MAG_daw}. %
The three workflows are all designed to analyze genome data, yet they are different in purpose and design.
Furthermore, we use Rangeland~\cite{lehmannFORCENextflowScalable2021} as a non-bioinformatics workflow analyzing satellite data.

\subsubsection{RNA-Seq}

RNA-Seq analyzes RNA sequencing data. 
RNA is a precursor to protein, and RNA-Seq analysis is essential for a better understanding of gene expression and regulation in a given context. %
RNA data consists of small sequences of RNA, which the workflow assembles before comparing the resulting RNA sequences to other conditions, e.g., healthy cells versus cells infected by a pathogen.
The RNA-Seq workflow first checks data quality by testing and filtering. 
Then, the RNA sequences are mapped against reference sequences to determine the number of genes expressed in a given condition. 
Post-processing tasks do quality controls and generate human-readable files.\looseness=-1 %

We use an RNA-Seq dataset of \textit{Homo sapiens}~\cite{rnaseq_human} for the experiments.
Additionally, we also executed the workflow with a second data set containing different lines of \textit{Drosophila melanogaster}~\cite{rnaseq_droso} to study the variance of peak memory demand of different input datasets in Section~\ref{subsec:task_memory_behavior}.

\subsubsection{Sarek}

Sarek performs variant calling, an important workflow in precision medicine. 
It analyzes sequenced genomic data to detect variants.
Variants represent segments of genetic sequences that deviate from a given reference.
For example, a single gene can have multiple variants that can affect its expression in different ways. 
In the case of cancer, knowing which variations are present can help the clinician choose the appropriate treatment for a patient.
The Sarek workflow consists of 45 steps, as shown in Table~\ref{tab:workflowCharacteristics}.
In the end, Sarek outputs annotated variants from the input data, quality analysis metrics, and a set of plots that can be interpreted by biologists.

The dataset we use to test this workflow stems from a study by \citeauthor{sarek_data}~\cite{sarek_data} in which they studied estrogen receptor mutations to target anti-cancer therapies better.

\subsubsection{MAG}
The MAG workflow performs assembly, binning, and annotation of metagenomes.
A metagenome is a set of all genomic data present in a given environment sample, such as the bacteria in our guts or in the soil.
Metagenomic analysis helps to understand and quantify the diversity of organisms in a given habitat.
For example, the analysis of water can help to understand whether it is contaminated by certain bacteria.
MAG takes long or short reads as input. 
After analysis, MAG returns quality control information, taxonomic classification, read assembly, and genome annotation.
The MAG workflow consists of 38 tasks, as shown in Table~\ref{tab:workflowCharacteristics}.

We use 17 samples of the human microbiome coming from a study on the influence of viruses on chronic diseases~\cite{mag_data}.
The samples were extracted from the gut, mouth, nose, skin, and vagina and can help the researchers understand bacteria-virus interactions and their correlation to diseases better.

\subsubsection{Rangeland}
The Rangeland workflow~\cite{lehmannFORCENextflowScalable2021} was initially developed to reassess the Rangeland degradation in the Mediterranean area relative to results reported 20 years ago~\cite{frantzRevisingThePast}.
Therefore, it uses the nowadays freely and publicly available satellite imagery from the Landsat archive\footnote{\url{https://www.usgs.gov/tools/earthexplorer}}.
The Rangeland workflow combines different processing steps of FORCE\footnote{\url{https://github.com/davidfrantz/force}} (Framework for Operational Radiometric Correction for Environmental monitoring~\cite{Frantz19a}).
Accordingly, it uses a completely different toolchain than the bioinformatics workflows discussed before.
The workflow consists of a pre-, a higher-level, and a post-processing stage.\looseness=-1

For this workflow, we use the same dataset as we used for the original study~\cite{lehmannFORCENextflowScalable2021,frantzRevisingThePast}.
The dataset comprises Landsat data collected between 1984 and 2006 with less than 70\% cloud coverage for Crete, Greece with a 30m spatial resolution.

%% file: figures/workflowInputs.tex
\begin{table}[!h]
    \centering
    \caption{Workflow characteristics}
    \begin{tabular}{l|r|r|r|r}
    
        \small
        \multirow{1}{*}{Workflow} & \multicolumn{1}{m{1cm}|}{\centering Abstract Tasks} & \multicolumn{1}{m{1cm}|}{\centering Physical Tasks} & \multicolumn{1}{m{1cm}|}{\centering Median Physical Tasks} & \multicolumn{1}{m{1cm}}{\centering Inputs} \\
        \hline
        RNA-Seq & 53 & 1,269 & 39.0 & 39 \\
        RNA-Seq (Drosophila) & 55 & 701 & 20.0 & 20 \\
        Sarek & 45 & 7,432 & 36.0 & 36 \\
        MAG & 38 & 7,618 & 17.0 & 17 \\
        Rangeland & 12 & 4,418 & 35.5 & 2,072 \\
        
    \end{tabular}
    \label{tab:workflowCharacteristics}
\end{table}

%% file: content/02.5_executionanalysis.tex
\subsection{Workflow Task Memory Behavior}
\label{subsec:task_memory_behavior}
For initial analysis of memory requirements within these workflows, we use the cluster setup as described in Section~\ref{sec:clusterSetup}, except for using the full memory available on the nodes.
We executed each workflow twice: once without using any memory sizing and once using the Witt-LR memory sizing approach.
This way, we also check whether all instances of all tasks work when they are limited to the predicted memory demand.
Next, we use the traces of our test runs to analyze the different memory usage patterns, the influence of the input data size on peak memory consumption, and how peak memory can vary for the same task between different runs.

\paragraph*{Workflow characteristics}
In Table~\ref{tab:workflowCharacteristics}, we present the general characteristics of all four workflows, where RNA-Seq was executed with two datasets. %
We see that for most workflows, the number of inputs influences the number of physical tasks generated by every abstract task (median physical tasks).
This is not the case for the Rangeland workflow due to its preprocessing: Rangeland maps the inputs to a small number of buckets and then continues processing the buckets.
Increasing the number of physical task instances allows online memory-sizing approaches to learn from more samples, resulting in better predictions for subsequent tasks.
The RNA-Seq workflow performs more abstract tasks for the \textit{Drosophila} dataset than for the Human dataset, as it needs to determine required domain-specific sequencing properties that we only had available for the Human dataset.

The large number of physical tasks shows that it is impractical to ask users for memory predictions for all task instances.

\paragraph*{Task patterns}\label{sec:taskPattern}
\input{figures/nftasks}
Next, we plot the size of each task's input data against the peak memory consumption for every abstract task.
We present four task patterns in Figure~\ref{fig::peakmemoryByInput}.
We draw a linear regression for the data points based on all samples of each task and the offset linear regression Witt-LR.
Moreover, we show the 95th percentile as another memory prediction approach~\cite{wittFeedbackBasedResourceAllocation2019}.\looseness=-1

The taxonomic data visualization in Figure~\ref{fig:mag:krona} shows a clear linear correlation between input size and memory requirements.
Still, due to the remaining variance, the Witt-LR approach underpredicts six out of 34 tasks, which will lead to failures.
The 95th percentile only underestimates two tasks but overpredicts the others by a large margin.

For RNA-Seq (see Figure~\ref{fig:rnaseq:qualimap}), Witt-LR underpredicts five, and the percentile approach underpredicts two out of 39 tasks.
Notably, the slope of the linear regression line for this task's data decreases, which is unexpected since more input usually leads to equal or more peak memory consumption. 
It seems that input size alone cannot explain peak memory consumption.
Linear regression is unable to deal with such logical inconsistencies.

Figure~\ref{fig:rangeland:preprocessing} shows two clearly visible and distinct point clouds for the preprocessing of the Rangeland workflow.
When trying to fit a linear regression, the result also has a negative slope and clearly badly matches the given measurements.
Here, Witt-LR underpredicts 144 of 2,072 tasks and the 95th percentile 104.
Tasks with lower peak memory are caused by satellite images containing water or clouds.
These features are not relevant for the analysis of vegetation and are thus discarded in the preprocessing stage.

Finally, the Sarek workflow (see Figure~\ref{fig:sarek:basecalibrator}) yields observations without any immediate relationship between input size and memory requirements.

The four plots indicate that (a) linear correlations are not always given and that (b) a fixed user prediction at the abstract task level will cause overheads for many, if not most, task instances.

\paragraph*{Memory usage}
\input{figures/memPerCore}
The requested memory and cores inherently limit workflow parallelism, as the sum of resources allocated to tasks on a cluster node cannot exceed the available memory or cores.
Hence, the more memory and cores are allocated to any individual task, the fewer other tasks can be run in parallel.
This is observable on a single cluster node as well as entire clusters.
Memory will be the limiting factor if tasks request more memory per core than a node can offer.
In Figure~\ref{fig:memPerCoreUserDefined}, the user-defined memory per core ratio is shown.
For example, if a task requests 4~GB and 2 cores, the ratio is 2~GB/Core.
Notably, for Sarek, 76.2\% of the tasks request 2~GB/Core or less, and all tasks request less than or equal to 6~GB/Core.
This means that Sarek executions cannot profit from memory sizing in a setup with at least 6~GB/Core.
In contrast, MAG requests up to 40~GB/Core.
The actual memory usage per core is shown in Figure~\ref{fig:memPerCoreUsed}.
Over all workflows, 99.0\% of all tasks use less than 2~GB/Core.
However, only 37.2\% of the tasks \emph{allocate} less than 2~GB/Core.

\begin{figure}[!t]
\begin{minipage}[t]{0.48\columnwidth}
    \centering
    \includegraphics[width=\linewidth]{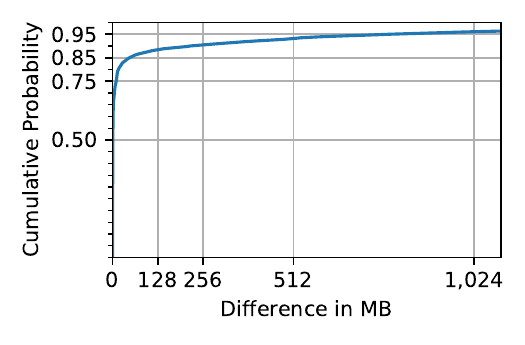}
    \caption{Cumulative distribution of peak memory difference between two runs for the same task in MB}
    \label{fig:peakMemDiff}
\end{minipage}
\hspace{0.02\columnwidth}
\begin{minipage}[t]{0.48\columnwidth}
    \centering
    \includegraphics[width=\linewidth]{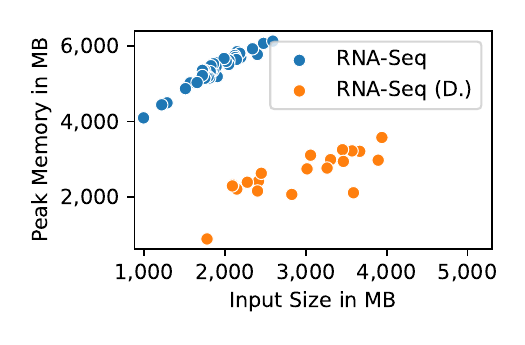}
    \caption{RNA-Seq coverage analysis for Human and Drosophila input data}
    \label{fig:rnaseqDiff}
\end{minipage}\hfill
\end{figure}

\paragraph*{Variance between runs}
Every workflow was executed two times and we compare the peak memory consumption for each task between both runs.
The peak memory might vary due to the multithreaded execution of many tasks.
As multiple tasks run on the same node and compete for I/O bandwidth, the available I/O bandwidth for individual tasks might differ between task executions.
This can cause delays of individual threads within a single task, resulting in a temporal shift of memory usage.
The cumulative distribution is shown in Figure~\ref{fig:peakMemDiff}.
54.3\% of all tasks vary by less than 1~MB between both executions and 85\% differ less than 48~MB.
However, 6.8\% differ by more than 512~MB, and the highest difference we observed was 5,707~MB.
The significant differences in peak memory requirements across repeated executions of the same tasks indicate that it is prudent to conservatively estimate task peak memory to avoid failures.

\paragraph*{Variance over different inputs}
Finally, we compare the behavior of the same task but with different input data.
For this, we tested the RNA-Seq workflow with a second data set.
Figure~\ref{fig:rnaseqDiff} shows the peak memory consumption for the coverage analysis of RNA-Seq when executed for Human or Drosophila data.
The human reference genome, which is larger than the one of Drosophila, leads to higher memory consumption.
This again indicates that a workflow developer has to overestimate the consumption if they are unaware of the actual data being used.
Moreover, it shows that hybrid approaches could suffer in performance when they pre-train on historical data.%

%% file: figures/nftasks.tex
\begin{figure}
    \centering
    
    \begin{subfigure}[t]{.5\columnwidth}
      \centering
      \includegraphics[width=0.95\linewidth,trim={5.4mm 5.6mm 4.89mm 5.28mm},clip]{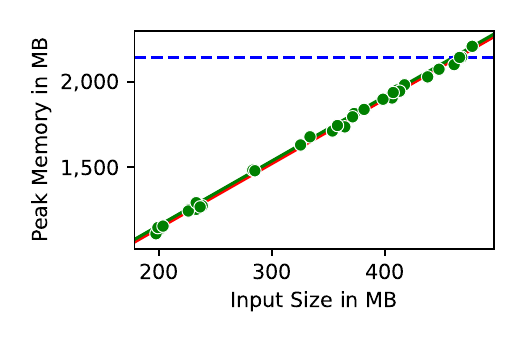}
      \subcaption{MAG: Taxonomic data visualization}
	 \label{fig:mag:krona}
    \end{subfigure}%
    \hfill
    \begin{subfigure}[t]{.5\columnwidth}
      \centering
      \includegraphics[width=0.95\linewidth,trim={5.4mm 5.6mm 4.89mm 5.28mm},clip]{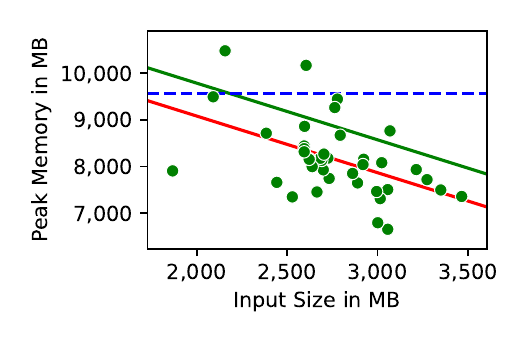}
      \subcaption{RNA-Seq: Quality control}
	 \label{fig:rnaseq:qualimap}
    \end{subfigure}%

    \medskip
    
    \begin{subfigure}[t]{.49\columnwidth}
      \centering
      \includegraphics[width=0.95\linewidth,trim={5.4mm 5.6mm 4.89mm 5.28mm},clip]{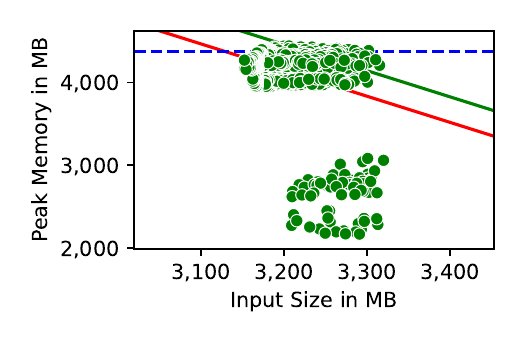}
      \subcaption{Rangeland: Preprocessing}
	 \label{fig:rangeland:preprocessing}
    \end{subfigure}%
    \hfill
    \begin{subfigure}[t]{.49\columnwidth}
      \centering
      \includegraphics[width=0.95\linewidth,trim={5.4mm 5.6mm 4.89mm 5.28mm},clip]{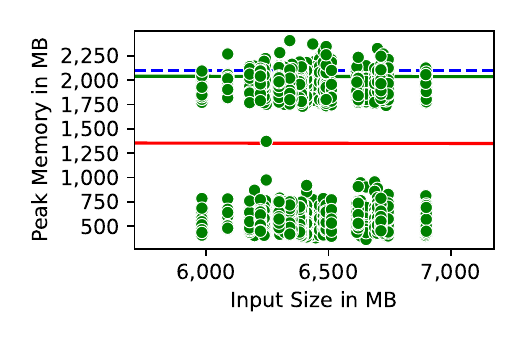}
      \subcaption{Sarek: Base quality score recalibration}
	 \label{fig:sarek:basecalibrator}
    \end{subfigure}%
    
    \medskip

    \begin{subfigure}{\columnwidth}
      \centering
      \includegraphics[width=\columnwidth,trim={0mm 0mm 0mm 0mm},clip]{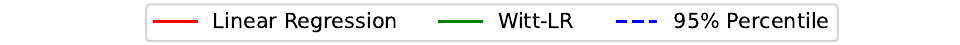}
    \end{subfigure}%
    
    \caption{Memory consumption depending on the input data for four exemplary tasks. In red, a linear regression, and in green, the regression shifted by the standard deviation between the predicted and the true value. The dashed blue line represents the 95th percentile.}
    \label{fig::peakmemoryByInput}
\end{figure}

%% file: figures/memPerCore.tex
\begin{figure}
    \centering
    
    \begin{subfigure}[t]{.5\columnwidth}
      \centering
      \includegraphics[width=0.850\linewidth,trim={4mm 4mm 3.8mm 4mm},clip]{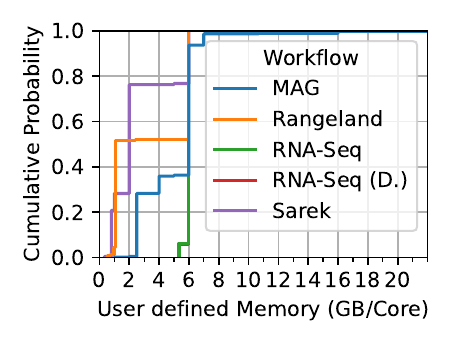}
      \subcaption{User-defined}
	 \label{fig:memPerCoreUserDefined}
    \end{subfigure}%
    \hfill
    \begin{subfigure}[t]{.5\columnwidth}
      \centering
      \includegraphics[width=0.850\linewidth,trim={4mm 4mm 3.8mm 4mm},clip]{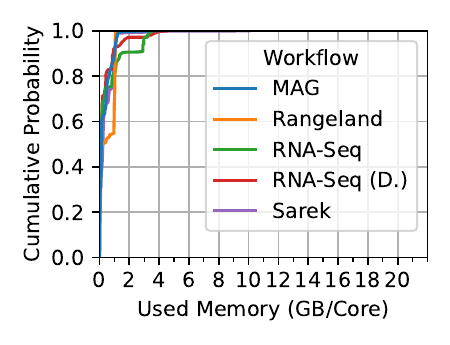}
      \subcaption{Used memory}
	 \label{fig:memPerCoreUsed}
    \end{subfigure}%
    
    \caption{Cumulative Distribution of user-defined and actual used memory per core assigned to a task. Probability of all physical tasks.}
    \label{fig::memPerCore}
\end{figure}

%% file: content/03_strategy.tex
\section{Approach}
In this section, we present our new strategy, Ponder, which we derived from our problem analysis.
First, we discuss key considerations for memory sizing approaches for real-world workflows.
Next, we explain our proposed strategy in detail.

\subsection{Key Strategy Considerations}\label{sec:keyStrategyConsiderations}
In Section~\ref{subsec:task_memory_behavior}, we have seen that a linear correlation between input sizes and memory requirements cannot always be assumed. 
In fact, only around half of all the tasks we tested show a positive Pearson correlation of $>0.3$ between input size and peak memory demand.
Linear models cannot reliably predict tasks that show no or only a weak correlation between input size and peak memory demand.
Additionally, we assume that there cannot be a negative linear relation between task size and peak memory.
Moreover, we found that when the same task is executed twice, the peak memory consumption can vary substantially, in some cases by more than 5~GB.
As the general sample variance scales by a factor of $\frac{1}{I}$, with $I$ being the increasing number of finished task instances, we expect the predictions to get continuously better in our online setting.\looseness=-1

\emph{Memory wastage} and \emph{Memory Allocation Quality (MAQ)} are two measures that have been proposed by related work to validate the performance of a memory sizing strategy \cite{tovarJobSizingStrategy2018a,wittFeedbackBasedResourceAllocation2019}.
Memory wastage~\cite{tovarJobSizingStrategy2018a} sums up the under-allocation wastage and over-allocation wastage. 
Under-allocation wastage occurs when a task fails due to insufficient allocated resources. 
It is defined by multiplying the requested memory by the runtime until the failure occurred.
Over-allocation wastage is the difference between requested memory and actual peak memory multiplied by the runtime for successful tasks.
MAQ sets the actual used memory time in relation to the requested memory time~\cite{wittFeedbackBasedResourceAllocation2019} as $\frac{U}{U+OW+UW}$, where $U$ is the used memory time, $OW$ the over-allocation wastage and $UW$ the under-allocation wastage.

We argue that both measures overestimate the importance of memory wastage.
Note that when optimizing for any of these two measures, systems must try to reduce over-prediction to a minimum. 
However, preferring extremely tight predictions increases the danger of task failures due to accidental underpredictions.
Moreover, many tasks are CPU- and not memory-bound.
Consider a node on which only one task can start because it will consume all available cores.
For this task, memory overprediction has no impact as long as the predictions do not exceed the available memory on the node.
In such cases, the memory ``wastage'' should be counted as zero, as the remaining memory cannot be allocated to other tasks anyway.\looseness=-1

Considering the high cost of task failures, our method primarily aims at reducing their likelihood while maximizing throughput.
To this end, we propose a rule-based heuristic that balances over-allocation and under-allocation wastage but, compared to previous work that was never tested in practice, is more lenient to overprediction when we cannot be sure about the true peak memory demands of a task.

\subsection{Our Memory Sizing Strategy}\label{sec:ourstrategy}
Ponder, our memory sizing strategy, addresses the requirements we identified by dynamically deciding between multiple methods.
It either applies linear regression based on memory consumption of the previously executed task instances, uses the maximum or minimal peak memory seen so far, or resorts to using the user-defined memory thresholds.

Our strategy is rule-based, adapts over time, and depends on the $I$ physical task instances that have finished at the time of prediction.
Each of these finished tasks gives us a tuple of ($x_i$,$y_i$), where $x_i$ was the input size of task $i$ and $y_i$ the peak memory consumption.
To predict a new task's peak memory $y^*_n$, we only make use of $x_n$ and the previously seen instances ($x_i$,$y_i$).
The algorithm is presented in Algorithm~\ref{alg:memPredicStrat}.
A short description follows below.

If fewer than five task instances have been observed and we, therefore, have only limited training data, we distinguish two cases depending on the input size $x_n$: 
\begin{enumerate}
    \item If $x_n$ is lower than the currently seen maximum input size, we set $y*_n$ to the highest peak memory consumption seen so far and add a static offset as a buffer.
    \item If $x_n$ is larger than any input size of tasks executed so far, we fall back to the user's default memory prediction $y_{user}$.
\end{enumerate}
We suggest an offset of 128~MB since 88.4\% of the tasks analyzed in Section~\ref{sec:problem_analysis} had a variance below this threshold.

If we saw five or more instances of the same abstract task, we calculate the Pearson correlation coefficient between the already seen physical task instances input sizes $X=\{x_1,...,x_{I}\}$ and their peak memory consumption $Y=\{y_1,...,y_{I}\}$ to determine whether to apply our linear regression predictor or not.
We distinguish two cases again:

\begin{enumerate}
    \item If the correlation coefficient is smaller $< 0.3$, no regression is applied: We return the highest value seen so far, again adding a buffer (and again suggesting 128~MB).
    \item Otherwise, we predict memory consumption using linear regression as follows below.
\end{enumerate}

As underprediction of memory consumption is much more costly in practice than overprediction, we introduce a weight term $\lambda$ as a hyperparameter, which penalizes an underprediction of memory more than an overprediction.
Formally, underprediction is defined as $y^*_n - y_n < 0$ for the predicted value $y^*_n$ and overprediction as $y^*_n - y_n > 0$ accordingly. 
Overpredictions are multiplied by the weight term $\lambda$. 
We empirically determined that $\lambda = \frac{1}{50}$ is a good fit for the nf-core workflows analyzed in Section~\ref{sec:problem_analysis}.
This tilts the linear regression towards overprediction, avoiding costly underpredictions, and offers a simpler alternative to the quantile regression presented by~\cite{wittLearningLowWastageMemory2019} where a model selection takes place.
The linear regression considers all $I$ finished tasks to minimize the loss.
Therefore, it takes $X$ and $Y$ as input for the regression model and the input size $x_n$ of the task for the specific prediction.
Predictions are represented as $\displaystyle f(X,Y,x) = a \cdot x + b$. %
We use an iterative optimizer to determine the best $a$ and $b$ for minimizing our loss function:
\begin{align*}
    \text{loss}(X,Y) &= \sum_{i=1}^I \text{error}\left(y_i,~a \cdot x_i + b\right) \text{with} \\
    \text{error}(y,f(x)) &= 
        \begin{cases}
            (y-f(x))^2,         & \text{if } y-f(x) > 0\\
            \lambda \cdot (y-f(x))^2, & \text{otherwise}
        \end{cases}
\end{align*}

After applying linear regression, we further safeguard our prediction based on the task instances we have seen with the following three checks:
\begin{enumerate}
    \item If we predict a smaller peak memory than previously observed, we use the smallest seen peak memory instead.
    \item If we predict a peak memory higher than previously observed, but a task with a larger input has finished successfully beforehand, we use the highest seen peak memory value instead.
    \item If $x_n$'s input size is larger than the input of all previously observed tasks and we predict a peak memory smaller than the highest peak memory observed, we also predict the highest seen peak memory.
\end{enumerate}

For the final prediction value $y^*_n$, we again add an offset as a buffer against underpredictions.
However, instead of using a static value (such as 128~MB) as before, we use the sample values to calculate a sample standard deviation similarly to~\cite{wittFeedbackBasedResourceAllocation2019} for the final prediction's offset.
Intuitively, larger sample standard deviations mean larger differences between the predicted regression line and our measured real memory consumption, which we reflect with a larger offset.
Compared to~\cite{wittFeedbackBasedResourceAllocation2019}, we use a weighted sample standard deviation instead of an unweighted one, i.e., if we have seen many $x_i$ values similar to our current input size $x_n$, the offset is more tilted to those values.
To calculate the weighted sample variance, we first calculate the weight $w_i$ of every observation, which decreases the further away the normalized input size is.
Then, using the weights $w_i$, the differences $d_i$ between predictions $f(x_i)$ and true memory consumption values $y_i$, and the weighted arithmetic mean of those differences $m$, we calculate the offset as the unbiased estimator of the sample standard deviation for a new $x_n$ as follows:

\begin{align*}
    &\text{offset}(X,Y,I) = 2 \cdot \sqrt{\frac{\sum_i^I w_i \cdot (d_i - m)^2}{v1-\frac{v2}{v1}}} \\
    &\text{with~} w_i = 1 - \frac{\lvert x_i - x_n \rvert}{\max_i^I(x_n,x_i)} + \frac{\max( 1 - \frac{I}{10}, 0)}{100}, \\
    &d_i = f(X,Y,x_i) - y_i,~ m = \frac{1}{v1} \cdot \sum_i^I d_i \cdot w_i, \\
    &v1 = \sum^I_i w_i \text{~and~} v2 = \sum^I_i w_i^2.
\end{align*}

We also multiply the calculated final prediction offset value by two, as this further decreases the likelihood of underpredictions.
In addition, we note that we add an extra term to the weights $w_i$ to ensure that all examples are weighted sufficiently when we have only seen fewer than ten examples.
In cases where the computed offset is below the configured static value (of 128~MB), we fall back to the static offset value instead.\looseness=-1

\begin{algorithm}
\caption{\small Ponder Strategy\\
\textbf{Input}: $X=\{x_1,...,x_{I}\}$, $Y=\{y_1,...,y_{I}\}$, $x_n$, $f(X,Y,x)$\\
\textbf{Output}: $y^*_n$\\
This algorithm predicts the peak memory $y^*_n$ for a physical task instance based on all finished physical task instances for the same abstract task.}
\begin{algorithmic}[1]
\label{alg:memPredicStrat}
\small
\IF{$I < 5$}
    \IF{$\max_i^I x_i > x_n$}
        \RETURN $\max_i^I y_i$ + 128~MB
    \ELSE
        \RETURN $y_{user}$
    \ENDIF
\ELSE
    \IF{Pearson($x_1 \dots x_I$,$y_1 \dots y_I$) $<$ 0.3}
        \RETURN $\max_i^I y_i$ + 128~MB
    \ELSE
        \STATE let $y^*_n = f(X,Y,x_n)$
        \IF{$y^*_n < \min_i^I y_i$}
            \STATE $y^*_n = \min_i^I y_i$
        \ELSIF{$y^*_n > \max_i^I y_i$ \textbf{and} $\max_i^I x_i > x_n$}
            \STATE $y^*_n = \max_i^I y_i$
        \ELSIF{$x_n > \max_i^I x_i > $ \textbf{and} $y^*_n < \max_i^I y_i$}
            \STATE  $y^*_n = \max_i^I y_i$
        \ENDIF
        \STATE $y^*_n = y^*_n + \max(\text{offset}(X, Y, I), 128~\text{MB})$
        \RETURN $y^*_n$
    \ENDIF
\ENDIF
\end{algorithmic}
\end{algorithm}

%% file: content/05_experiments.tex
\section{Evaluation}
In this section, we will first briefly introduce our prototype.
Next, to evaluate our approach and prototype, we perform different experiments using the workflows presented in Section~\ref{sec:workflows}.
We compare our memory sizing strategy Ponder to Witt-LR and the user memory allocation.
Furthermore, we evaluate the memory sizing strategies with different scheduling strategies, assessing how the sizing influences scheduling and how scheduling influences the sizing.
Both our prototype and our measurements are available on Github\footref{git:cws}\footref{git:experiments}.

\input{content/04_prototype}

\subsection{Memory Sizing Strategies}
We use three different memory sizing strategies:
\begin{enumerate}
    \item \emph{User}: The User strategy does not apply any memory sizing but uses user-defined values. 
    These values are configured in the workflow definition and have been set by the workflow developers.
    \item \emph{Witt-LR}: We choose Witt-LR~\cite{wittFeedbackBasedResourceAllocation2019} as a state-of-the-art memory sizing baseline.
    It performs linear regression to predict peak memory consumption based on a task's input size.
    We choose the Witt-LR variant, which adds the sample standard deviation as an offset to improve robustness.
    \item \emph{Ponder}: Our Ponder strategy is described in Section~\ref{sec:ourstrategy}. 
\end{enumerate}

We use the user-defined values for a second attempt at running a task with both sizing strategies (Witt-LR and Ponder) if the first attempt fails.
This is a change from the original failure strategy presented by Witt-LR~\cite{wittFeedbackBasedResourceAllocation2019} and is used because the user-defined memory sizes are usually conservative and, thus, serve as a good fallback.

\subsection{Scheduling}
Nextflow submits tasks as soon as they are ready to run to Kubernetes, which then schedules tasks to nodes.
For this scheduling, we evaluate four different strategies, two of them in two versions.
\begin{itemize}
    \item\emph{Original:}
        The first scheduling strategy is Kubernetes's default strategy, which is commonly applied in practice when no other method has been configured.
        Here, Kubernetes assigns tasks using FIFO (First In - First Out) with gap filling.
        If no resources for the next task to schedule are available, it looks for the first task that could be started on the remaining available resources.
    \item\emph{Rank:}
        The second strategy prioritizes tasks based on their rank, and in the case of a tie, it prefers tasks with larger input sizes.
        The rank is the number of tasks on the longest path to an end task.
        Consider Figure~\ref{fig:exampleAbstractDAG}: for task A, the rank is 3, and for task C, the rank is 1.
        Tasks with larger inputs usually take longer, and starting them earlier reduces the risk of waiting for them when all other tasks have finished and a workflow joins.
    \item\emph{LFF (Min):}
        The Least Finished First strategy was proposed by \citeauthor{wittFeedbackBasedResourceAllocation2019}~\cite{wittFeedbackBasedResourceAllocation2019}.
        The strategy prefers tasks that belong to an abstract task, where the least physical tasks have finished.
        The idea is to generate samples for prediction faster.
        In the case of ties, LFF prefers tasks with smaller inputs.
        Tasks with smaller inputs tend to finish faster, and thus, samples are collected faster.
    \item\emph{LFF (Max):}
        We also use the LFF approach with another tiebreaker, preferring tasks with larger inputs.
        The idea is to quickly obtain an upper bound for the memory prediction for large tasks, leading to fewer costly prediction errors.
    \item\emph{Generate Samples (Min):}
        The GS approach is a hybrid of Rank and LFF.
        We prioritize tasks with less than five finished physical task instances.
        In case of a tie, we use the rank but prefer tasks with a smaller input size to generate samples faster.
        If we have five or more finished instances, we prioritize tasks based on the Rank strategy.
    \item\emph{Generate Samples (Max):}
        This strategy is the same as before, but we use Rank also for ties when we saw less than five instances.
\end{itemize}

\subsection{Setup}\label{sec:clusterSetup}
We execute our experiments in a cluster with eight nodes, all equipped with an Intel Xeon Silver Processor 4314, which has 16 cores and 32 threads.
Each node offers 256~GB of memory and contributes two 3.5~TB Nvme SSDs to a Ceph cluster.
The nodes are managed by Kubernetes 1.27.7, with the feature \emph{InPlacePodVerticalScaling} enabled.
We reduce the usable memory to 96~GB per node, which brings the memory-to-core ratio to 3~GB/core.
We decided on this ratio, since it makes all workflows memory-limited (compare Figure~\ref{fig:memPerCoreUserDefined}) to show the advantage of improved memory sizing.

To execute the workflows, we use Nextflow in version 23.10.1.
For nf-core, we manually set the maximal memory to 64~GB.
We apply the same upper bound to our memory sizer.
Moreover, we set the minimal memory to predict 128~MB to prevent the pod from being unable to be started due to insufficient memory.
Both the lower and the upper limit are applied to all memory sizing approaches.

\subsection{Results}
\input{figures/resultsFigure}
For our experiments, we executed 12 strategy combinations for four workflows. 
This gives us a total of 48 combinations, 20 each for Witt-LR and Ponder.
Every combination is executed three times, and we consider the execution with median makespan in this evaluation.
We split the evaluation into two parts.
First, we compare the performance of our Ponder approach against Witt-LR when the same scheduling strategy is used.
Second, we compare the influence of different scheduling strategies.

\paragraph{Influence of the sizing strategy}
First, we compare the makespan for the same scheduling strategy.
Makespan is shown in Figure~\ref{fig:makespan}.
For all but one workflow-scheduling combination, Ponder is better and reduces the makespan on average by 21.8\%.
The highest makespan reduction took place for Rangeland using LFF(Min).
Here, Ponder's makespan is 43.5\% lower than the according execution for Witt-LR.
One reason is the many underpredictions for the preprocessing tasks, as indicated in Figure~\ref{fig:rangeland:preprocessing}.
LFF(Min) Witt-LR failed for 991 (32.4\%) of the preprocessing tasks, which is significantly more than we expected in Section~\ref{sec:taskPattern}, showing the difficulties when predictions are performed on a limited set of tasks.
Ponder, on the other hand, only failed for 25 tasks (1.2\%).
Also, for the other workflows, the increased makespan can be explained by Witt-LR's much higher failure rate compared to our strategy.
Our Ponder strategy has, on average, 93.8\% fewer task failures (Figure~\ref{fig:failedTasks}).
This is due to the fact that Ponder waits for more samples before it predicts and uses a higher offset.
Only 7.3\% of the predictions made by the Ponder strategy are smaller than the ones made by Witt-LR.
However, in this case, the predicted task fails in 3.7\% for Ponder and in 1.1\% of the cases for Witt-LR.
While significantly fewer tasks fail for Ponder, it is notable that they also fail quicker (Figure~\ref{fig:ttf}).
52.4\% of the tasks fail before half of their runtime for Ponder, and only 23.9\% do so for Witt-LR.

The high failure rate, together with our failure strategy, is also reflected in the MAQ measures (as explained in Section~\ref{sec:keyStrategyConsiderations}) in Figure~\ref{fig:maq}.
Here, on average, Ponder achieves a 71.0\% higher MAQ than Witt-LR.
An exception is the MAG workflow, where no significantly higher MAQ is achieved. 
Ponder achieves a higher MAQ than Witt-LR for 17 out of 20 workflow-scheduling combinations.
At the same time, our Ponder strategy also reduces allocated CPU and memory time by 13.3\% and 30.7\%, respectively.
While the increased CPU time is related to Witt-LR's frequent failures, the increased memory time is also caused by the failure strategy that uses user-defined values.
However, the Ponder strategy's average CPU usage in the cluster is higher, on average, by 12.7\%.
Ponder uses the cluster's CPUs by 46.7\% on average (Figure~\ref{fig:clusterUsageCPU}).
This indicates that we have fewer situations where a workflow joins and waits for a small number of tasks.

Figure~\ref{fig:diffPredictedAndUsed} shows the difference between the predicted value and the actual peak memory consumption.
Notably, the Ponder strategy has higher overprediction, leading to fewer task failures.

\paragraph{Influence of the scheduling strategy}
Next, we investigate the influence of the scheduling strategy.
Our Ponder strategy has a lower makespan than the Original and Rank strategy without sizing in 19 cases, while Witt-LR outperforms the Original in 16 cases and Rank in 12 cases.
This shows that using memory sizing is not always better.
An unfortunate scheduling and workflow combination can increase makespan.
On average, Ponder achieves the lowest makespan with GS(Max) and Witt-LR with Rank.
For both sizing strategies, LFF(Max) performed better than LFF(Min) for all workflows except Rangeland for Ponder and MAG for Witt-LR. 
This is not obvious as LFF(Min), on the one hand, can train models faster but schedules tasks with larger input sizes later.
Scheduling tasks later that could potentially take longer can create situations in which the workflow cannot continue until one task has finished.
The smaller average number of task instances per abstract task explains why LFF(Min) performed better than LFF(Max) for MAG and shows that fast training is important.
We can see that the scheduling strategy has more influence using the Witt-LR strategy than Ponder.
For Ponder, the makespan standard deviation is 15.3~min on average, and for Witt-LR, 25.7~min.\looseness-1

For Ponder, GS(Max) achieves the highest average MAQ (0.54) and the worst for Rank (0.51).
In contrast, Witt-LR's highest average MAQ (0.35) is with Rank, but the value is significantly lower than that of Ponder.

For Ponder, the success of sizing is almost independent of the scheduling strategy.
Success rates fluctuate by less than 1.3\%.
For Witt-LR, however, the success rate differs by 5.1\% between the scheduling strategies.
Depending on the scheduling strategy, Ponder sizes 1.8\% fewer tasks and Witt-LR 3.1\%.

Figure~\ref{fig:ttf} shows the relation between time to failure and the time to succeed. 
Notably, we can only calculate the time to failure when a task fails.
Accordingly, we have fewer observations for Ponder.
If Ponder fails, it achieves a much quicker time to failure than Witt-LR on average.
This is the case when we underpredict the peak memory by larger margins.

%% file: content/04_prototype.tex
\subsection{Prototype}
We base our prototype on our Common Workflow Scheduler Interface (CWSI), which connects SWMSs to resource managers (RMs)~\cite{lehmannHowWorkflowEngines2023}.
The interface allows the submission of tasks with workflow-specific information that is otherwise not available to resource managers.
Moreover, the workflow DAG is transferred to the resource manager as it unfolds, which enables complex workflow-aware scheduling strategies.

We extend the Common Workflow Scheduler implementation available for Kubernetes with memory sizing and a simple interface to implement memory sizing strategies for Nextflow workflows.
We see this as an advantage over studies that implement their strategies in simulation tools, which require additional implementation steps to make the strategies available to real users.

Implementing the memory sizing component in the scheduler of the RM, rather than the scheduler of the SWMS, allows tasks to be submitted to the RM as soon as they are ready to run and still be sized later, without the sizing becoming a dependency or complicated update routines.
For any sizing strategy, a user can define a lower and upper bound. 
Predictions will not exceed these values.

As we decided for Nextflow as the SWMS, we adapted the nf-cws plugin\footref{git:nfcws} to transfer required information using the extended CWSI.

Our changes have been merged into the main branches of both projects and are thus available to Nextflow users.

%% file: figures/resultsFigure.tex
\begin{figure*}[!t]
    \centering
        \caption*{Legend for the strategies}
    \includegraphics[width=\linewidth]{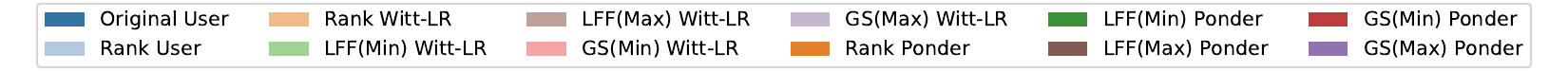}
\begin{minipage}[t]{0.66\linewidth}
    \vspace{0pt}
    \begin{subfigure}[t]{0.48\linewidth}
        \centering
        \includegraphics[width=\linewidth,trim={4mm 4.2mm 3.8mm 3.8mm},clip]{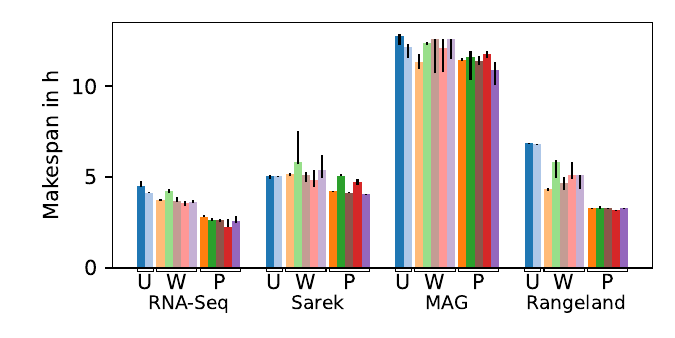}
        \caption{Makespan}
        \label{fig:makespan}
    \end{subfigure}
    \hfill
    \begin{subfigure}[t]{0.48\linewidth}
        \centering
        \includegraphics[width=\linewidth,trim={4mm 4.2mm 3.8mm 3.8mm},clip]{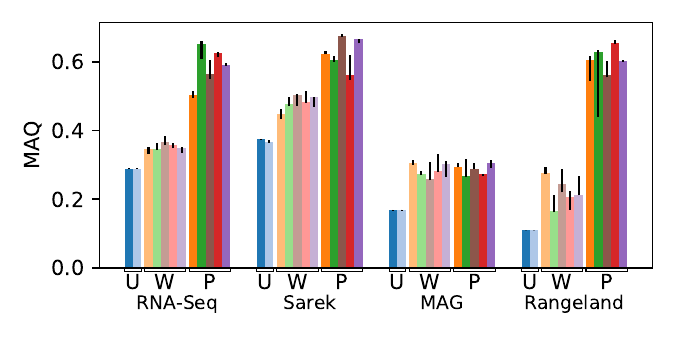}
        \caption{MAQ}
        \label{fig:maq}
    \end{subfigure}
    \hfill
    \begin{subfigure}[t]{0.48\linewidth}
        \centering
        \includegraphics[width=\linewidth,trim={4mm 4.2mm 3.8mm 3.8mm},clip]{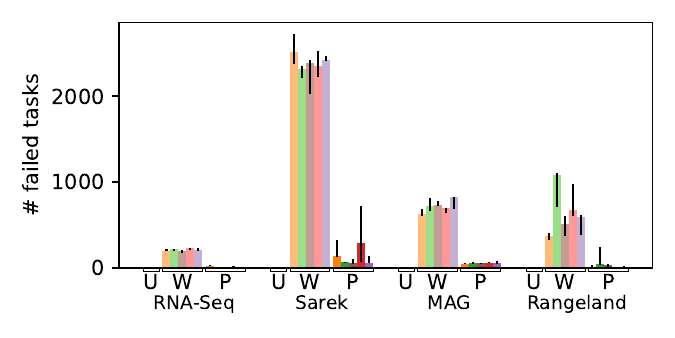}
        \caption{Failed tasks}
        \label{fig:failedTasks}
    \end{subfigure}
    \hfill
    \begin{subfigure}[t]{0.48\linewidth}
        \centering
        \includegraphics[width=\linewidth,trim={4mm 4.2mm 3.8mm 3.8mm},clip]{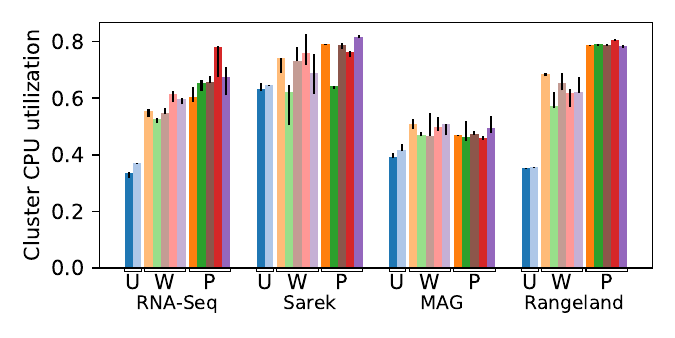}
        \caption{Cluster CPU utilization}
        \label{fig:clusterUsageCPU}
    \end{subfigure}
    \caption{Median metrics with variance over three workflow executions. The bars are grouped by workflows and by the sizing strategy. The single character indicates whether the sizing strategy is \textbf{U}ser, \textbf{W}itt-LR, or \textbf{P}onder.}
    \label{fig:combined}
\end{minipage}
\hfill
    \begin{minipage}[t]{0.33\linewidth}
        \vspace{0pt}
        \nextfloat
        \begin{subfigure}[t]{\linewidth}
            \centering
            \includegraphics[width=\linewidth,trim={1.5mm -2.0mm 0.0mm 3.8mm},clip]{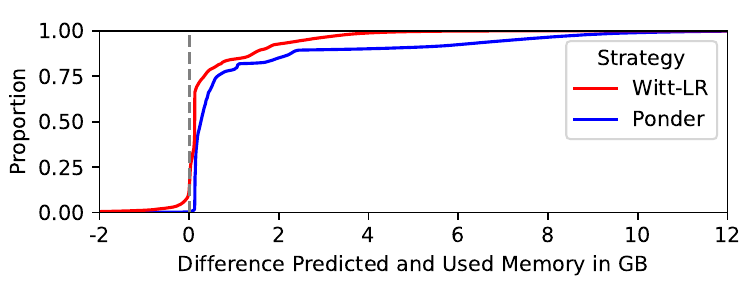}
            \caption{Difference between the predicted and actual peak memory by prediction strategy}
            \label{fig:diffPredictedAndUsed}
            \vspace{3mm}
        \end{subfigure}
        \begin{subfigure}[t]{\linewidth}
            \centering
            \includegraphics[width=\linewidth,trim={1.5mm 16.8mm 0.0mm 3.8mm},clip]{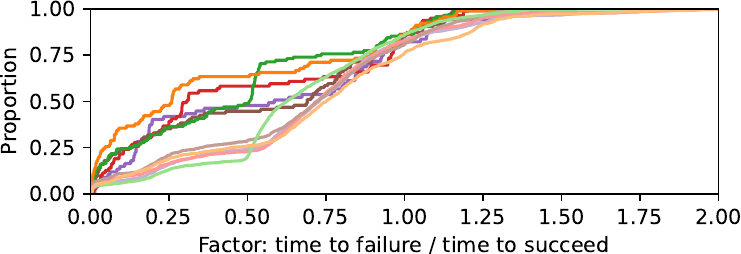}
            \caption{Time to failure divided by success time}
            \label{fig:ttf}
        \end{subfigure}
        \caption{Cumulated distribution for all sizing strategies for time to failure and predicted memory in relation to the actual values for successful execution.}
        \label{fig:factor}
    \end{minipage}%
\end{figure*}%

%% file: content/06_relatedwork.tex
\section{Related Work}

Our previous work~\cite{wittFeedbackBasedResourceAllocation2019} proposes an online method for combining scheduling and memory prediction.
For memory prediction, it uses a percentile predictor and a linear regression model that depends on a task's input size.
We use this as our baseline \emph{Witt-LR} in this paper.
Among the investigated scheduling policies, we previously observed a tradeoff between MAQ and makespan that was also influenced by different failure strategies.
In contrast to Witt-LR, Ponder uses linear regression with unequal loss and switches dynamically between regression and minimum or maximum values.
Moreover, we use a variable plus a fixed offset instead of only the standard deviation.
In our evaluation setup based on real workflow executions,  we also evaluate all approaches without a synthetic, fixed time-to-failure ratio.\looseness=-1

In~\cite{wittLearningLowWastageMemory2019}, we develop a prediction model that minimizes memory wastage instead of the prediction error.
Again, the task input size is used as the model input while a quantile regression is applied.
We tested multiple failure-handling strategies and showed that this choice has a big impact on the overall performance.
Instead of quantile regression, with Ponder, we use an unequal loss to shift the regression function.
Moreover, we add a more selective offset.

\citeauthor{tovarJobSizingStrategy2018a}~\cite{tovarJobSizingStrategy2018a} propose a job sizing strategy that aims to maximize job throughput and minimize job memory wastage.
Their method considers the probability distribution of peak memory usage to determine the actually assigned amount of memory.
The method does not incorporate the size of the input task; rather, it minimizes the sum of the probabilities of memory peaks where the memory is greater than the allocated memory value.

In~\cite{baderLeveragingReinforcementLearning2022}, we test two different reinforcement learning (RL) methods to minimize CPU and memory wastage.
The reward function of the reinforcement learning methods implicitly discourages the agents from underestimating resources.
Our RL methods can be pre-trained on historical executions and applied in an online manner, as a hybrid approach.
Neither of our RL methods considers task input sizes and is, thus, prone to producing inaccurate predictions for tasks with fluctuating resource requirements.

In~\cite{bader2023predicting}, we develop an online method that can be pre-trained on historical executions, predicts a task's memory consumption over time, and dynamically adjusts the task limits.
The method predicts a task's runtime, segments the predicted runtime, and predicts the peak memory for each segment.
Both predictions are adjusted by an offset to avoid underpredictions.
While this is more fine granular, it requires time series monitoring.
Further, the resource manager needs to be able to change the memory allocation during runtime and a scheduler that deals with memory time series, features that are not widely supported.\looseness=-1

\citeauthor{tovarDynamicTaskShaping2022}~\cite{tovarDynamicTaskShaping2022} present a method that splits the tasks into subtasks, matching a specific memory requirement.
This avoids the memory prediction problem and instead predicts the task size.
Furthermore, in the event of task failure, the memory is not increased; instead, the task is split again.
This method is not applicable to the black-box scientific workflow tasks as we consider them in this paper.

The interplay of resource demand predictions and scheduling was also addressed in other areas of research, such as cloud-based applications~\cite{kumar2019comprehensive} or in Big Data frameworks~\cite{sayehBlink22,myungMLBasedMemoryPrediction2021}.
However, neither of these is directly applicable to scientific workflows in clusters; the latter mismatches with the black-box model of scientific workflow systems, while the former assumes elastic environments.

%% file: content/99_conclusion.tex
\section{Conclusion}
In this paper, we analyzed the peak memory demand of four nf-core workflow tasks.
We show that state-of-the-art memory wastage measures are not ideal for task memory sizing.
Based on our analysis, we created a new peak memory sizing strategy, Ponder.
Our experimental evaluation shows that Ponder consistently reduces a workflow's makespan compared to the state-of-the-art baseline, resulting in an average of 71.0\% higher Memory Allocation Quality.
In addition, we showed that the baseline method also has a much higher failure rate when executed online than when applied to the full data set.
Furthermore, the effectiveness of our Ponder strategy is less dependent on the specific scheduling strategy used.

In future work, we would like to further intertwine memory prediction and scheduling.
More specifically, we plan to explore scheduling tasks so that we strategically generate new data points for prediction models.
Moreover, we plan to also consider a node's remaining memory when choosing offsets, in addition to input data size.